\documentclass [11pt]{article}
\usepackage{booktabs}		
\usepackage{multirow}
\usepackage[pdftex]{graphicx}
\usepackage[dvips]{epsfig}
\usepackage{amsmath}
\usepackage{amssymb}
\usepackage{color}
\usepackage{authblk}

\begin{document} 

\title{\LARGE{Human mobility networks and persistence of rapidly mutating pathogens}}
\author[1,$\dagger$]{Alberto Aleta}
\author[2,$\dagger$,\footnote{\textit{Present address}: 91, Av du G\'en\'eral Leclerc 92340, Bourg la Reine, France.}]{Andreia N. S. Hisi}
\author[1,4]{Sandro Meloni}
\author[2,$\ddagger$]{Chiara Poletto}
\author[2,3]{Vittoria Colizza}
\author[1,3,4]{Yamir Moreno}

\affil[1]{\small{Institute for Biocomputation and Physics of Complex Systems, University of Zaragoza, Zaragoza, Spain}}
\affil[2]{\small{Sorbonne Universit\'es, UPMC Univ Paris 06, INSERM, Institut Pierre Louis d$'$\'Epid\'emiologie et de Sant\'e Publique (IPLESP UMRS 1136), Paris, France}}
\affil[3]{\small{ISI Foundation, Turin, Italy}}
\affil[4]{\small{Department of Theoretical Physics, University of Zaragoza, Zaragoza, Spain}}
\affil[$\dagger$]{\small{equal contributors}}
\affil[$\ddagger$]{\small{corresponding author chiara.poletto@inserm.fr}}

\date{}
\maketitle

\section*{}
Rapidly mutating pathogens may be able to persist in the population and reach an endemic equilibrium by escaping hostsÕ acquired immunity. For such diseases, multiple biological, environmental and population-level mechanisms determine the dynamics of the outbreak, including pathogen's epidemiological traits (e.g. transmissibility, infectious period and duration of immunity), seasonality, interaction with other circulating strains and hostsÕ mixing and spatial fragmentation. Here, we study a susceptible-infected-recovered-susceptible model on a metapopulation where individuals are distributed in subpopulations connected via a network of mobility flows. Through extensive numerical simulations, we explore the phase space of pathogen's persistence and map the dynamical regimes of the pathogen following emergence. Our results show that spatial fragmentation and mobility play a key role in the persistence of the disease whose maximum is reached at intermediate mobility values. We describe the occurrence of different phenomena including local extinction and emergence of epidemic waves, and assess the conditions for large scale spreading. Findings are highlighted in reference to previous works and to real scenarios. Our work uncovers the crucial role of hosts' mobility on the ecological dynamics of rapidly mutating pathogens, opening the path for further studies on disease ecology in the presence of a complex and heterogeneous environment.

\newpage
\clearpage

\section{Introduction}
Many pathogens are able to persist in a host population by repeatedly reinfecting individuals who lose the immunity acquired during infection~\cite{matt_j._keeling_modeling_2008}. This mechanism is particularly important for the case of rapidly mutating pathogens when mutation is associated to antigenic drift -- e.g. for influenza virus~\cite{zhang_co-circulation_2013} and respiratory syncytial virus~\cite{weber_modeling_2001}. The probability of pathogen persistence is then determined by the interplay between the transmissibility and infection time scales (i.e. duration of infection and immunity period), along with ecological factors like seasonality, interaction with other circulating pathogens, mixing patterns among individuals, their spatial distribution and mobility.

Waning of immunity and partial immunisation have been accounted for by several modelling studies~\cite{pease_evolutionary_1987,gomes_infection_2004,bansal_impact_2012,truscott_essential_2011,nasell_influence_2013,clancy_effect_2009} addressing detailed immunological mechanisms~\cite{yuan_evolutionary_2013}, interaction with other strains~\cite{truscott_essential_2011,andreasen_dynamics_1997}, human contact structure~\cite{bansal_impact_2012}, environmental factors~\cite{shaman_absolute_2010}, seasonality and temporal variation in transmission~\cite{dushoff_dynamical_2004,li_evolutionary_2015,casagrandi_sirc_2006}. The fate of an emerging pathogen in terms of persistence/extinction was also studied, both as a theoretical problem~\cite{nasell_influence_2013,clancy_effect_2009} and through applications to the emergence of a novel influenza strain~\cite{zhang_co-circulation_2013,asaduzzaman_coexistence_2015}. However, the vast majority of studies on waning of immunity and persistence have disregarded spatial structure so far. 

Recently, it has been shown that the geographical distribution of the host population and the rate of traveling among distant locations determine the time-scale of human mixing, hence affecting the probability of having large-scale epidemic spread~\cite{balcan_phase_2011,belik_natural_2011,meloni_modeling_2011,colizza_epidemic_2008}  and the incidence profile~\cite{poletto_host_2013,poletto_characterising_2015,grenfell_travelling_2001,marguta_impact_2015}. For the case of childhood diseases (e.g. measles) previous studies showed that these dynamical effects may impact disease persistence and alter the critical community size~\cite{bartlett_measles_1957,bolker_space_1995,bolker_impact_1996,keeling_metapopulation_2000,rozhnova_impact_2014,jesse_divide_2011,hagenaars_spatial_2004}, i.e., the population threshold above which a disease is able to persist \cite{bartlett_measles_1957}. Results reported so far however suggest that the persistence probability is determined by the interplay between the different spatial ingredients -- e.g. distance and frequency of travels, environmental influences on transmission --  in a non-trivial way. Indeed, different contrasting hypotheses have been suggested, with persistence maximised for intermediate levels of spatial coupling~\cite{keeling_metapopulation_2000,jesse_divide_2011}, for maximum levels of coupling~\cite{hagenaars_spatial_2004}, or either of the two depending of the level of spatial heterogeneity in transmission~\cite{rozhnova_impact_2014}. 

The spatial structure of human populations could have important and non-trivial effects on the spreading of diseases also when waning of immunity is at work, affecting the possible dynamical regimes following an outbreak. In this paper, we address this problem by means of a stochastic discrete spatially-explicit metapopulation model. We reconstruct the dynamics of an emerging infection through extensive numerical simulations. We characterise the roles played by  the spatial fragmentation, the coupling, and the structure of the mobility network through which individuals travels from one subpopulation to another. We focus on the regime of parameters that best describes influenza-like infections. Our metapopulation framework  realistically reproduces the statistical properties of the human population, i.e.,  strong heterogeneities and correlations in the population distribution and mobility. We perform a detailed analysis of the mechanisms shaping the probability of pathogen persistence and the spatial dynamics following emergence in the short and long term. 
Our approach offers the theoretical and computational framework to further explore the effects of other features not taken into account here (e.g., multiple strains, seasonality, and others) in order to gain a full understanding of pathogen's persistence and be able to perform realistic data-driven simulations.

\section{Methods}

\begin{figure}%
\includegraphics[width=1\linewidth]{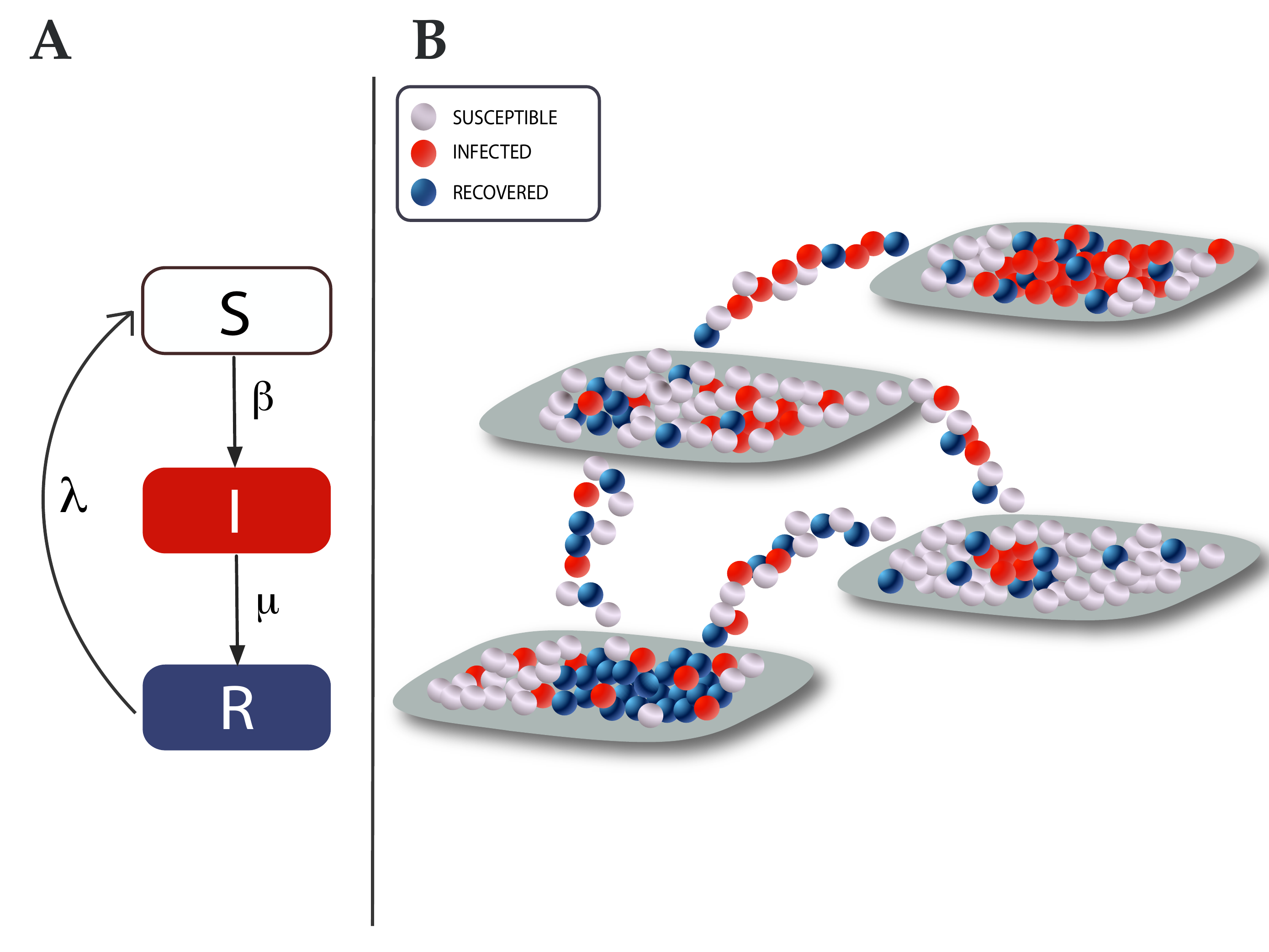}%
\caption{Schematic representation of the stochastic metapopulation model. Panel (a) represents the disease dynamics that follows a SIRS scheme, where individuals change states with given transition probabilities (Susceptible (S), Infected (I), Recovered (R) and back to S when the acquired immunity is lost). Panel (b) illustrates the structure of the metapopulation network, which consists of subpopulations that are connected by individuals traveling from one to another through a mobility network.}%
\label{fig1}%
\end{figure}

We use a stochastic individual-based metapopulation model \cite{levins_demographic_1969,hanski_ecology_2004,lloyd_spatial_1996}, where mobility and infection of individuals are explicitly accounted for as discrete-time processes. Individuals mix homogeneously within the local communities (also called subpopulations, patches or nodes of the metapopulation system), whereas at the global level the coupling between these subpopulations is introduced by individuals that travel along mobility connections (Figure~\ref{fig1}). The model is thus represented in terms of a network of links, i.e. mobility flows, connecting different nodes representing local populations. 

Massive data and extensive research studies have uncovered the statistical properties of the distribution of human populations and their associated mobility, pointing out their network structure, characterised by large variabilities and non trivial correlations~\cite{balcan_multiscale_2009,simini_universal_2012,chowell_scaling_2003,barrat_architecture_2004}. In particular, the population of urban areas and the number of connections between locations (considering e.g. air-transportation or commuting) span several orders of magnitude and more populated centres are also often the hubs of the mobility network. We account for these properties by assuming a power law distribution for the patches' connectivity and a linear relationship between connectivity and population. Specifically, we considered a network of $V$ patches, each of them (node $i$ of the network) characterised by a population of size $N_i$, and a degree $k_i$ that  represents the number of subpopulations through which node $i$ is linked via mobility connections. The nodes' degree distribution is assumed to be of the form $P(k)\sim k^{-\gamma}$ $-$ henceforth we have set $\gamma=3.0$ and disregarded eventual topological correlations for simplicity. Mobility fluxes are assigned to each link by assuming that in each subpopulation, individuals may travel to neighbouring subpopulations with a probability $p$ per unit  time. Departing individuals from node $i$ choose at random one of the available $k_i$ links, so that the probability of traveling along a connection is given by $p/k_i$. This process yields a population distribution that at equilibrium is proportional to the number of connections, namely, $N_i= k_i \langle N \rangle/ \langle k \rangle$ -- where $\langle N \rangle$ and $\langle k \rangle$ are the average population per node  and average degree of the metapopulation network, respectively~\cite{colizza_epidemic_2008}. 

The infection dynamics is implemented through the Susceptible-Infected-Recovered-Susceptible compartmental scheme~\cite{pease_evolutionary_1987}. The evolution of the dynamics results from the following transition rules iterated at each time step, corresponding to one day: 
\begin{itemize}
\item [$(i)$] Susceptible individuals within each subpopulation $i$ may be infected with probability $1-\left(1-\frac{\beta}{N_i}\right)^{I_i}$, where $\beta$ represents the transmission probability, and $N_i$ and $I_i$ denote, respectively, the total population and the number of infected individuals in node $i$;
\item[$(ii)$] Infected individuals transition with probability $\mu$ to the recovered compartment, that represents a state in which individuals are  immune to  the disease~--~here immunity is assumed to be complete; 
\item[$(iii)$] eventually, recovered individuals lose their immunity and get back to the susceptible state. This occurs with probability $\lambda$, that is the inverse of the average immunity period, $L$. 
\end{itemize}

The previous set of transition rules disregards vital dynamics. This choice simplifies the dynamical characterization of the model and it is motivated by our interest in diseases with durations of both infection and immunity periods much shorter than the average life duration (e.g. influenza). Starting with a fully susceptible population, 
the disease is seeded in a randomly chosen node by infecting 0.5\% of its population. The disease may then propagate inside the local population and spread to neighbouring populations due to hosts movements. Progression of the disease and host movements are stochastic events that are numerically modelled through multinomial processes. For each scenario considered,  $10^3$ independent epidemics are simulated starting from different seeded subpopulations. 

We consider influenza-like infections spreading in space as a prototypical example for our study. We fix the duration of infection $\mu^{-1}=2.5$ days and the basic reproductive number $R_0=\beta/\mu=2.0$. The population is assumed to be composed of $10^{8}$ individuals divided in $V$ patches. We explored scenarios characterised by: (i) different spatial fragmentation of the host population, considering $V=10^2,\,10^3,\,10^4$; (ii) a varying coupling between patches expressed by host mobility, considering $p \in [10^{-6}, 1]$; (iii) different topological structures of mobility connections, comparing the heterogeneous network described by a power-law degree distribution to a homogeneous Erd\H{o}s-R\'{e}nyi graph~\cite{erdos_random_1959} described by a Poisson degree distribution with the same average value; (iv) a varying immunity period, considering $L \in [1,800]$ days.

For each set of parameters, we numerically monitor the global prevalence (i.e. the total fraction of infectious individuals) and the number of infected patches $D(t)$ in time. We also estimate the probability that the pathogen persists in the population at the global level, $P_{gl}$, by computing the fraction of stochastic runs for which the epidemic reaches the endemic state,  once a given set of parameter values is considered. The same quantity can be defined at the local level to measure the probability of local persistence $P_{loc}$. This can be computed considering a closed population, or a population integrated into a metapopulation framework thus open to incoming and outgoing fluxes of individuals.
At the local level, we also define an outbreak as a single epidemic (infecting at least $0.1\%$ of the population) that continuously persist in the population. 
Repeated outbreaks instead are considered as distinct waves following local extinction. 
These behaviours are characterised for the different scenarios explored, in order to provide a coherent picture of the mechanisms for virus persistence and endemic equilibrium in the spatially structured population.

\section{Results and Discussion}

In the following, we present first the results regarding the persistence of the pathogen in the population and the mechanisms behind it, followed by a detailed characterisation of the spatial dynamics across the scenarios explored.

\subsection{Persistence}
\label{subsec:persistence}

\subsubsection{Interplay between waning of immunity and mobility}
\begin{figure}%
\begin{center}
\includegraphics[width=0.8\linewidth]{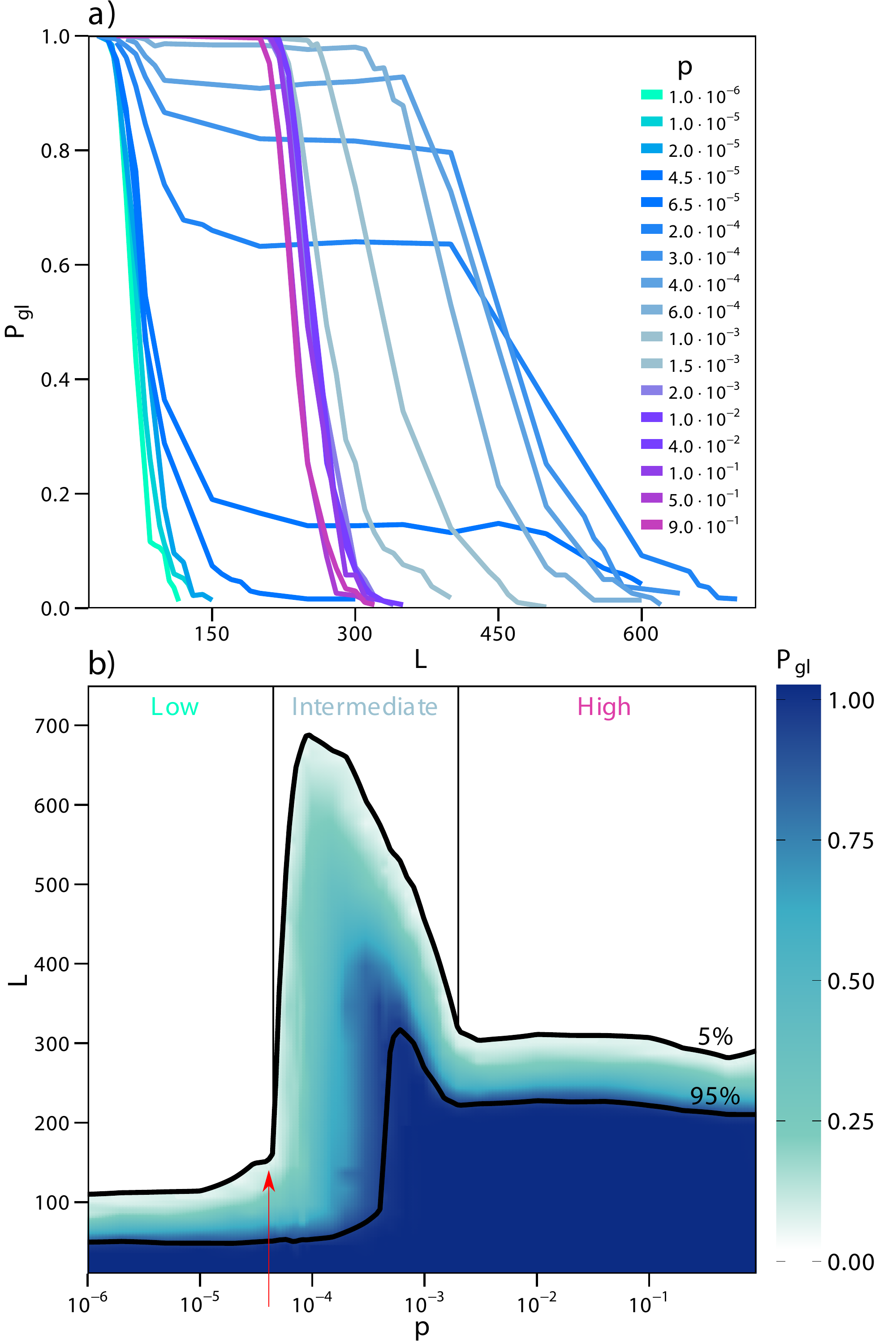}%
\end{center}
\caption{
$(a)$ Persistence probability $P_{gl}$ as a function of the duration of immunity $L$ for several values of the traveling probability $p$. Three different regions can be distinguished: low mobility $10^{-6} \leq p \leq 4.5 \cdot 10^{-5}$ (cyan), intermediate values $4.5 \cdot 10^{-5} \leq p \leq 10^{-3}$ (light blue) and high mobility $2\cdot10^{-3} \leq p \leq 1$ (purple). $(b)$ Color map of $P_{gl}$ in the $(p,L)$ plane. Black curves highlight the contours with $P_{gl}=5\%$ and $P_{gl}=95\%$ as indicated in the figure. The red arrow indicates the value of the global invasion threshold for the mobility network and disease parameters here considered, $\frac{1}{\langle  N \rangle} \frac{\langle k
 \rangle^2}{\langle k^2\rangle - \langle k \rangle} \;\frac{ \mu R_0^2}{ 2 (R_0-1)^2}= 4.5 \cdot 10^{-5}$. Both panels show results obtained with $V= 10^4$, $\langle N \rangle=10^4$ and power law degree distribution. }%
\label{fig:persistence_main}%
\end{figure}

The probability of persistence is found to be very high and equal to 1 for short immunity periods; once immunity becomes longer, a reduction of the probability of persistence is observed (Figure~\ref{fig:persistence_main}$a$). This behaviour is common to all host mobility rates explored. When the waning of immunity is very fast, there is a high rate of renewal of susceptible individuals that can be newly infected by the pathogen, thus ensuring its survival in the population. A slower rate of renewal makes persistence less likely and the specific behaviour of the decrease in persistence probability depends on the coupling between patches. Reduction may occur through a drop to zero persistence (e.g. $p=10^{-6},10^{-5},etc.$ and $p=10^{-2},\,10^{-1},\,etc.$), or a drop to an intermediate positive value of persistence probability that is maintained for a large range of immunity period, before reaching $P_{gl}=0$ (e.g. $p=10^{-4},2\cdot 10^{-4},etc.$).

The dependence of persistence drop on mobility is highlighted in Figure~\ref{fig:persistence_main}$b$. Our results show that it is possible to distinguish three mobility regimes~--~namely, low mobility (red), intermediate mobility (brown), and high mobility (green). These regimes are approximately delimited by $p< 4.5\cdot10^{-5}$, $p\in [4.5\cdot10^{-5},2.0\cdot 10^{-3}]$, and $p > 2.0\cdot 10^{-3}$ (referring to the profile of 5\% persistence probability) respectively.

In the low mobility regime, persistence at the global level is mainly determined by the probability that the pathogen would persist locally in the seeded population.  If $p$ is low enough so that no infectious individual travels out of the infected patch during the first wave of the epidemic, pathogen survival following the first wave depends on the length of the immunity period. A long enough immunity period prevents the replenishment of healthy individuals in the short term, thus leading to the extinction of the epidemic in the seeded patch, with no chances for global spread. If the availability of new susceptible hosts is instead provided  fast enough  by a short immunity period, the epidemic can survive the first wave in the seeded population, reaching a local endemic equilibrium. This, on its turn, can slowly seed the epidemic in neighbouring patches even when small mobility rates are considered, thanks to the local equilibrium condition. This same process is then occurring in each newly seeded patch, so that propagation  unfolds at the global level and a high probability of persistence is observed. 

The condition distinguishing between an epidemic trapped locally in a patch and one that spreads at the global level has been quantified in previous works in terms of the global invasion threshold~\cite{cross_duelling_2005,ball_epidemics_1997,balcan_phase_2011,poletto_human_2013,colizza_epidemic_2008,colizza_invasion_2007,apolloni_metapopulation_2014}.  Its analytical expression, provided in~\cite{colizza_epidemic_2008}, provides an upper bound for the low mobility regime: 
\begin{equation}
p  < \frac{1}{\langle N \rangle} \frac{\langle k \rangle^2}{\langle k^2\rangle - \langle k \rangle} \;\frac{ \mu R_0^2}{ 2 (R_0-1)^2}.
\label{eq:gl_th} 
\end{equation}
Once informed with the demographic and topological features of our spatially structured population, this condition is found to provide a good approximation to the limit of the low mobility regime (red arrow in Figure~\ref{fig:persistence_main}$b$).

The spatial structure of the population becomes less important once the system is found in the high mobility regime. In this case the probability of moving from one patch to another is so high as to largely increase the opportunity of mixing of the populations belonging to different patches. The system behaves effectively as a single population of the size of the full metapopulation (i.e. in our case $10^8$ individuals) affected by a SIRS dynamics. The conditions for global persistence are therefore those described before for the seeded population in the low mobility regime. Differently from before, here local persistence is equivalent to global persistence, due to the large degree of mixing across space. The values of the immunity period at which a given persistence probability is observed (e.g. $L\simeq300$ for $P_{gl}=5\%$) differ from the case of low mobility regime ($L\simeq100$), because of the difference in the population size~--~i.e., a single patch in the low mobility regime and effectively the full metapopulation size in the high mobility regime. 

The intermediate mobility regime displays a higher persistence probability for a given duration of the immunity period, compared to low and high mobility regimes (Figure~\ref{fig:persistence_main}$b$). The contour line at a fixed $P_{gl}$ (e.g. $P_{gl}=5\%$ or $95\%$) exhibits a maximum for intermediate values of the traveling probability $p$ that is reached for high values of $L$. In addition, a smoother variation of the persistence probability is observed by varying $L$ for a given $p$, compared to the faster transitions recovered in the low and high mobility regimes. The intermediate degree of mixing provided by intermediate mobility seems to enhance stochastic effects responsible for local extinction, thus having an important impact on the persistence at the global level.

Maximum persistence for intermediate level of coupling was also reported in the context of childhood diseases, where SIR models with continuous external introduction of the infection were considered~\cite{keeling_metapopulation_2000,jesse_divide_2011}. No consensus was however found in this context, with other studies reporting different findings. For instance, Hagenaars \emph{et al.} found an increase in the persistence with increasing levels of the coupling among subpopulations~\cite{hagenaars_spatial_2004}. Rozhnova \emph{et al.} confirmed this finding when homogeneous transmission across patches is considered, and additionally showed that heterogeneity in transmission among patches selects dynamical regimes with different spatial effects~\cite{rozhnova_impact_2014}. Different results obtained from different models show how the detailed mechanism of susceptible replenishment, together with infectious transmission and recovery, and spatial mixing, have an impact on the resulting persistence. Prompted by this, we assess in the following the role played by the spatial fragmentation of the population and by the mobility network on our model where the replenishment of individuals is obtained through  immunity waning.

\subsubsection{Role of spatial fragmentation and mobility network}

\begin{figure}%
\begin{center}
\includegraphics[width=0.8\linewidth]{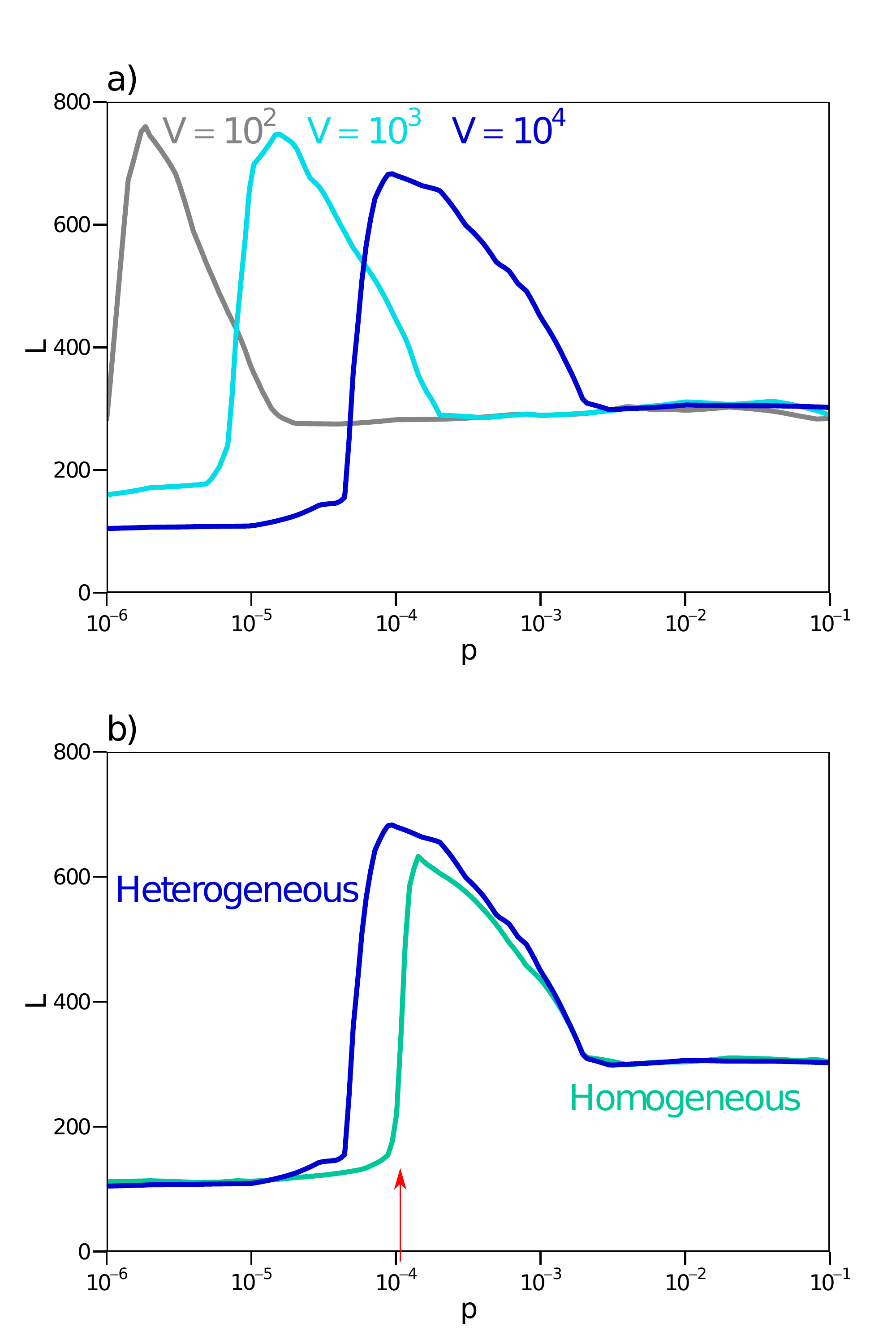}%
\end{center}
\caption{
$(a)$ Fragmentation analysis. Curves represent $P_{gl}=5\%$ for three values for the number of patches $V= 10^2, 10^3 \mbox{ and } 10^4$ subpopulations. In all the cases the total population has been kept constant $N_{tot}= 10^8$ ending up with an average population per subpopulation of $\langle N \rangle=10^6, 10^5 \mbox{ and }10^4$ respectively. Other parameters as in Fig.~\ref{fig:persistence_main}. $(b)$ Impact of network topology. Curves represent $P_{gl}=5\%$ for homogenous (Erd\"os-R\'enyi graph~\cite{erdos_random_1959}, green line) and heterogenous (power low distribution, blue line) mobility networks.  Both networks have the same number of patches $V=10^4$ and similar average degree $\langle k \rangle$. Other parameters as in Fig.~\ref{fig:persistence_main}.}%
\label{fig:persistence_size}%
\end{figure}

Spatial fragmentation is explored by subdividing the total population of $N=10^8$ hosts into $V= 10^2, \; 10^3, \; 10^4$ patches. Global persistence in the three scenarios shows the same qualitative behaviour described in the previous section (Figure~\ref{fig:persistence_size}$a$). The separation between different mobility regimes however depends on the number of patches considered.  More precisely, it is a demographic effect due to a varying  average population size per patch. Referring to the transition from low mobility to intermediate mobility regime, we have indeed that the estimate expressed in Eq.~(\ref{eq:gl_th}) depends on $\langle N \rangle$. Increasing the number of patches from $V= 10^2$ to $V=10^3, \; 10^4$, the average size of the population in a given patch decreases of one and two orders of magnitude, respectively, thus inducing a corresponding increase in the mobility value of the transition from low to intermediate regimes. As a result, the maximum of the global persistence is obtained for correspondingly increasing values of the traveling probability.

It is interesting to note that the behaviour observed in the high mobility regime is the same across the three scenarios. This confirms that the resulting dynamics when $p$ is high enough corresponds effectively to the local persistence dynamics of a single population of $10^8$ individuals, as discussed in the previous subsection. 

If the topological structure connecting the patches via mobility links is altered and homogenised, we find that higher values of mobility need to be reached for the transition from low to intermediate mobility to occur (Figure~\ref{fig:persistence_size}$b$).  As expected, heterogeneous mobility patterns (as the ones adopted in Figure~\ref{fig:persistence_main}) favour hosts mobility resulting in a larger global persistence for lower traveling probabilities. This is consistent with what predicted by the global invasion of Eq.~(\ref{eq:gl_th}) as shown by the arrow in Figure~\ref{fig:persistence_size}$b$.  At high mobility regimes, the structure connecting the links does not play any role, and global persistence is the same. Mixing is high enough so that the structured population behaves effectively as a single population regardless of the topology through which such mixing occurs.

\subsection{Spatial dynamics}
\begin{figure}%
\begin{center}
\includegraphics[width=1\columnwidth]{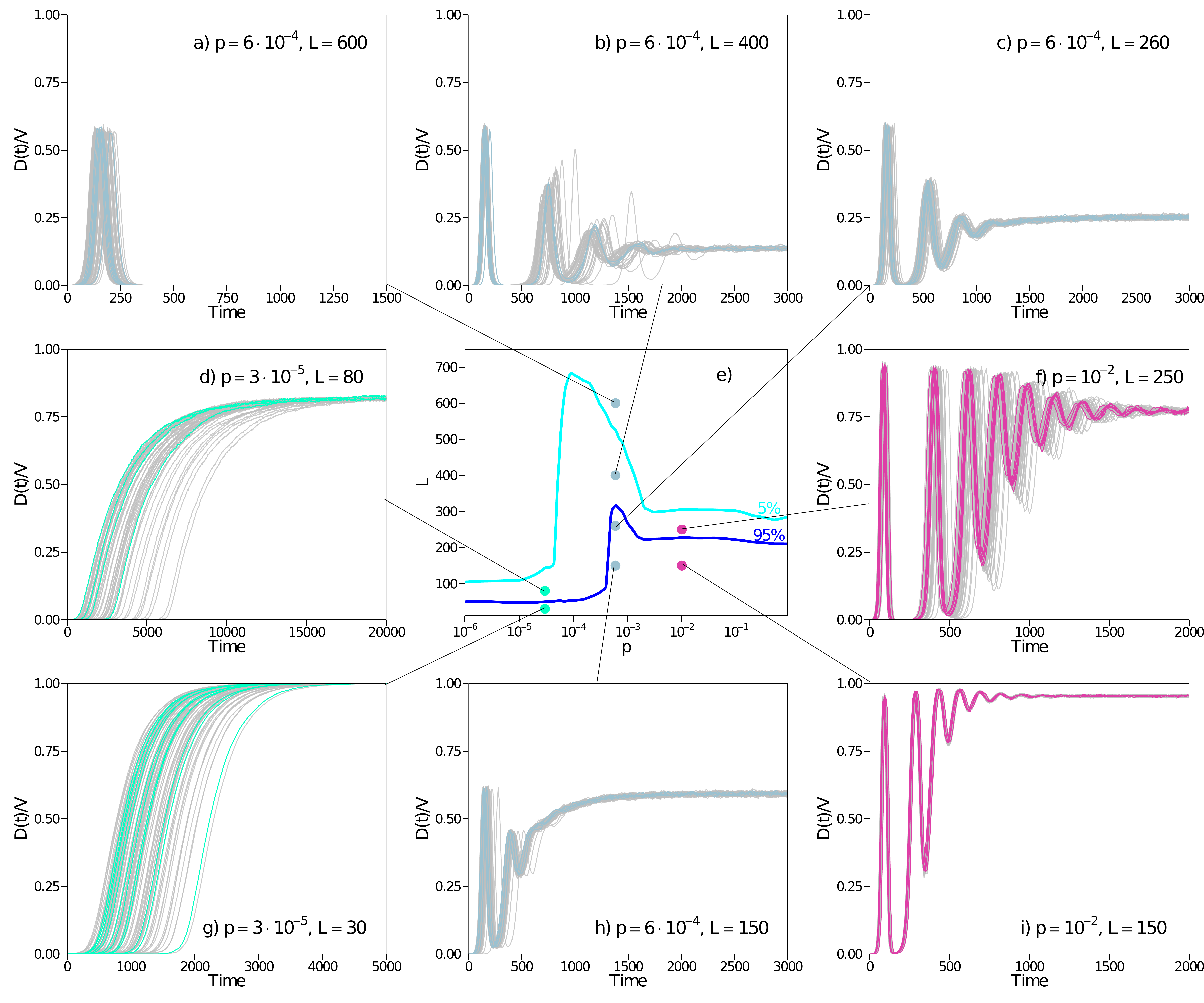}%
\end{center}
\caption{
Spatial dynamics. The external panels show the time evolution of the fraction of infected subpopulations $D(t)/V$ for different values of the duration of immunity $L$ and traveling probability $p$. The central panel shows the different regions where  each value of $L$ and $p$ has been chosen. In each panel curves corresponding to $10^2$ runs with different initial conditions are showed. To facilitate visualisation 90\% of them are smoothed. }%
\label{fig:dynamics}%
\end{figure}

In this section we go more in depth into the understanding of the spatial dynamics following the emergence of a pathogen in the fully susceptible and spatially structured population. Starting from the global persistence diagram reported in Figure~\ref{fig:persistence_main}, we characterise the epidemic diffusion in space considering a single value of reference for $p$ in each of the three mobility regimes ($p=3\cdot10^{-5}$ in the low, $p=6\cdot10^{-4}$ in the intermediate, and $p=10^{-2}$ in the high mobility regime), associated with different values of the immunity period $L$ (Figure~\ref{fig:dynamics}).

\subsubsection{Low mobility coupling}
We first analyze the low mobility regime, by characterizing the spatial epidemic dynamics obtained for $p=3\cdot10^{-5}$ and $L=30$ leading to $P_{gl}=1.0$ (panel $g$ of Figure~\ref{fig:dynamics}), and $p=3\cdot10^{-5}$ and $L=80$ leading to $P_{gl}=0.50$ (panel $d$). As discussed previously, global persistence is determined by local persistence of the disease at the patch level. The very high persistence observed for $p=3\cdot10^{-5}$ and $L=30$ is ensured by an endemic equilibrium that is reached in all patches after a given time, as signalled by 100\% of patches experiencing an outbreak (i.e. $D(t)/V=1$ after a given $t$). Even with a low probability of moving from one patch to another, the initially seeded population is able to transfer infected hosts to neighbouring populations in the long time limit thanks to the endemic equilibrium reached. On their turn, newly infected patches can further propagate to the infection reaching all patches of the system, thus achieving a high probability of persistence of the pathogen at the global level, fuelled by each endemic dynamics. 

\begin{figure}%
\begin{center}
\includegraphics[width=\linewidth]{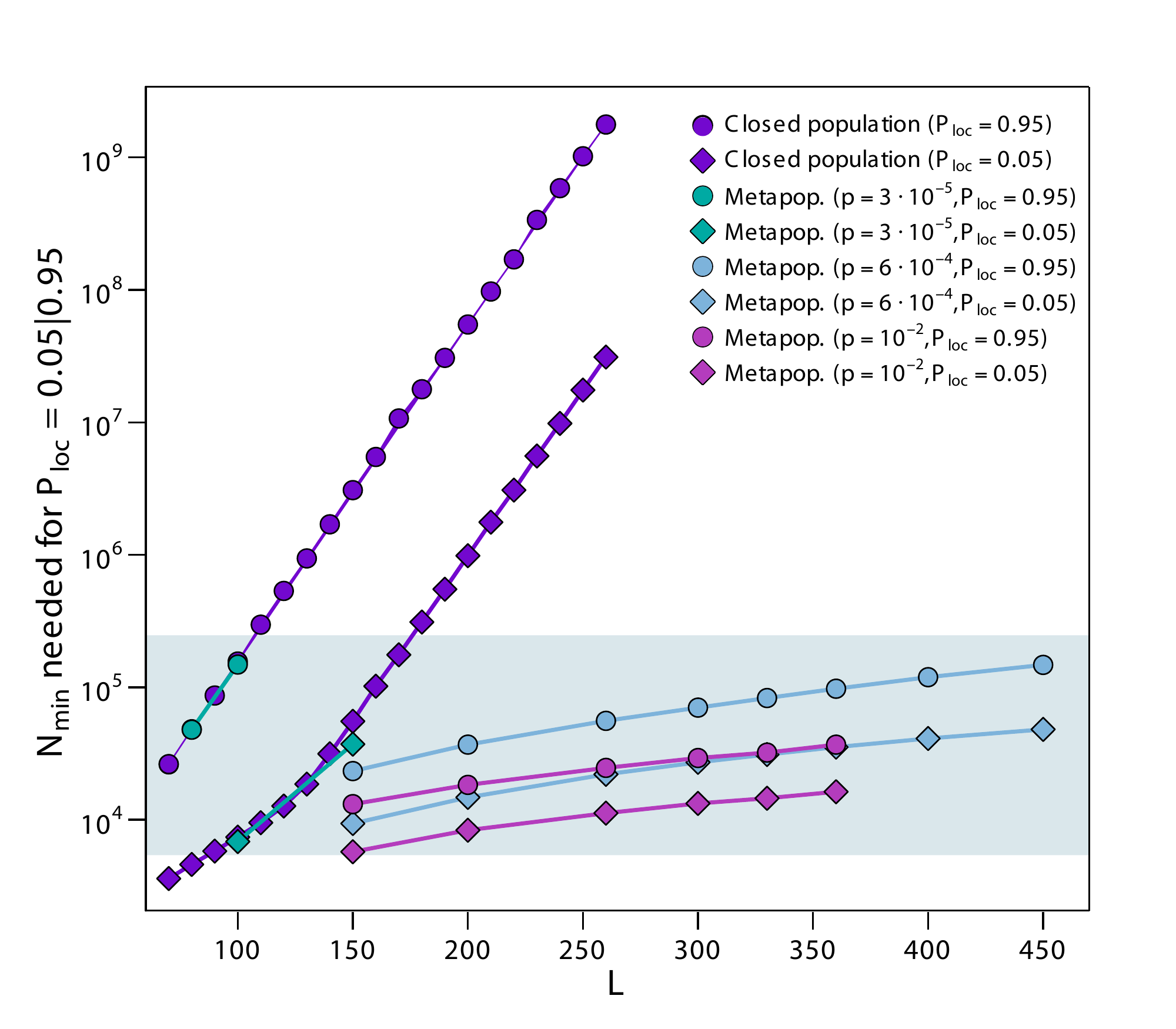}%
\caption{Minimum population size $N_{min}$ needed to achieve a local persistence probability $P_{loc}=5\%$ and $P_{loc}=95\%$ as function of the duration of immunity $L$ for the case of an isolated population and for populations of the metapopulation system. The shaded gray area denotes the possible population values in our system.}%
\label{fig:persist_pop}%
\end{center}
\end{figure}

The same traveling probability  leads to lower $P_{gl}$ once the immunity period increases, from $L=30$ to $L=80$ (panel $d$). Here a similar mechanism of local persistence and spatial diffusion ensured by the endemic equilibrium is at play, but it does not occur in all patches. A the end of our simulations we find indeed that only a fraction of patches is experiencing an outbreak ($D/V\simeq 80\%$). The different dynamics occurring across patches depends on the patch population size. For a fixed value of the immunity period $L$, there exist a minimum population $N_{min}$ needed in order to achieve a certain probability of local persistence (Figure~\ref{fig:persist_pop})~\cite{bartlett_measles_1957,nasell_influence_2013,wearing_estimating_2009,brooks-pollock_herd_2009}. The population size of the patch of our system varies approximately in the range $[5\cdot 10^3, 2.7 \cdot 10^5]$, due to the heterogeneous topology connecting the patches. If $L=30$, we find that all population values are above the minimum population needed to achieve local persistence and mobility can act as spatial spreader to reach global persistence. If $L=80$, we find that a portion of the patches do not have enough population to achieve local persistence. In particular, to obtain a local persistence probability higher than 95\% the minimum population size is $5\cdot 10^4$. The epidemic is able to spread at the global scale, however smaller peripheral patches do not meet the condition for local endemic equilibrium. This results in a slower spatial spread, compared to the case of $L=30$ (as displayed by the time axis in panels $d$ and $g$ of Figure~\ref{fig:dynamics}), and an epidemic that is not able to endemically reach all subpopulations ($D/V<1$).

\begin{figure}%
\includegraphics[width=\textwidth]{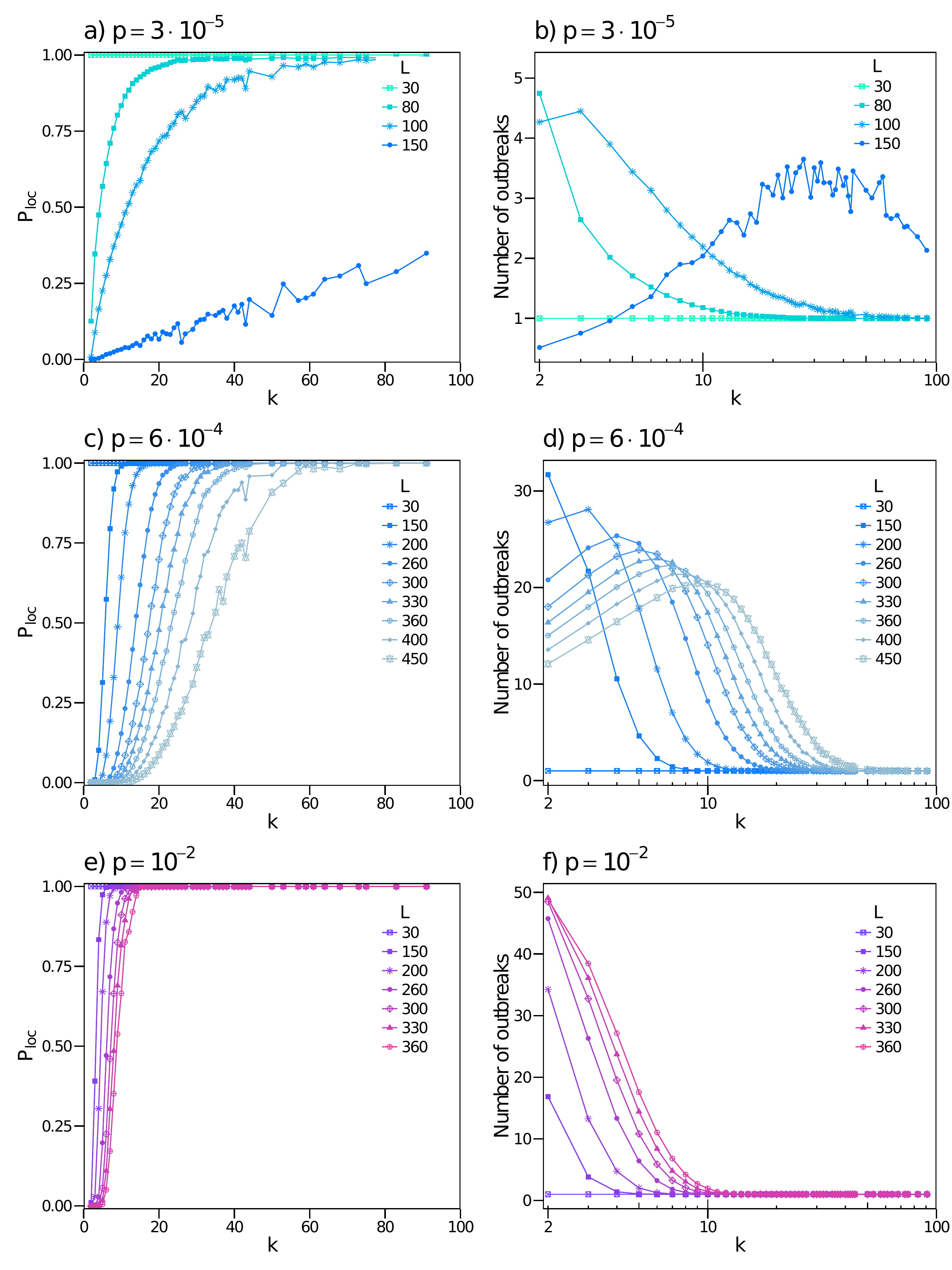}%
\caption{ $(a)$, $(c)$, $(e)$ Probability of local persistence in patches with degree $k$ for different values of the immunity period $L$ at $p=3 \cdot 10^{-5}$ (top row), $6 \cdot 10^{-4}$ (central row)  and  $10^{-2}$ (bottom row). The probability is computed as the fraction of outbreaks occurring in a node of degree $k$ lasting at least $2,000$ time-steps. $(b)$, $(d)$, $(f)$ Average number of distinct outbreaks experienced during a simulation by a node of degree $k$ for different values of the immunity period $L$ at $p=3 \cdot 10^{-5}$ (top row), $6 \cdot 10^{-4}$ (central row)  and  $10^{-2}$ (bottom row). Statistics is limited to runs where global persistence occurs. All  measurements have been performed after a transient time of 2000 steps to focus on the endemic dynamics.}%
\label{fig:degree-block_analysis}%
\end{figure}

The dynamics for the $L=80$ case reflects a spatial hierarchy similar to the one studied in the context of childhood diseases -- notably, for measles epidemics in the United Kingdom~\cite{grenfell_travelling_2001,grenfell_cities_1998,marguta_impact_2015}. At  equilibrium, highly connected and populated patches are steadily infected, while the peripheral ones experience repeated outbreaks in the rare occasions in which infectious travellers arrive from the hubs and find enough susceptible individuals to induce a local outbreak. 
To characterise this hierarchy, we analyse how local persistence changes across classes of patches and quantify the number of outbreaks each class experience throughout the duration of the simulated epidemic. Classes of patches are defined by their connectivity, i.e. we focus on nodes having the same number $k$ of neighbouring subpopulations. Due to the correlation between degree of a patch and its population size, these connectivity classes correspond effectively to population classes. Increasing degree leads therefore to higher local persistence, because of the relation to the population size (Figure~\ref{fig:degree-block_analysis}$a$ for the low mobility regime). $P_{loc}$ reaches 1 in high degree nodes for small enough values of the immunity period (i.e. up to $L=100$ days among the cases shown in the panel). In addition, we find that in these conditions hubs consistently experience only one outbreak (Figure~\ref{fig:degree-block_analysis}$b$), corresponding to an epidemic that persists in the population after the first seeding event. Low-degree nodes instead experience repeated outbreaks seeded from neighbouring nodes interspersed with local extinctions. For higher values of $L$ (e.g. $L=150$), the maximum local persistence achieved even in highly connected nodes is smaller than one and global persistence is approximately $P_{gl}=0.03$. Our results show that when disease persistence occurs, it is driven by repeated outbreaks taking place in different locations. Patches with intermediate degrees have the highest turnover of consecutive epidemics (about 3 outbreaks measured in the explored timeframe, see Figure~\ref{fig:degree-block_analysis}$b$). Hubs present fewer outbreaks, as these last longer because of their larger population size and due to the continuous reseeding from neighbouring patches that lead to multiple consecutive waves. Peripheral nodes instead have less outbreaks because, being less reachable, they are not frequently reseeded after local extinctions by other nodes.

\subsubsection{Intermediate mobility coupling}
For intermediate values of the traveling probability $p$, the spatial dynamics displays multiple waves followed by an endemic behaviour, if the probability of global persistence is larger than 5\% (Figure~\ref{fig:dynamics}$b$, $c$ and $h$). For smaller persistence (e.g. $P_{gl}=0.01$, panel $a$), the epidemic goes extinct after an initial wave. Interestingly, the amplitude of the initial wave is determined only by the mobility, regardless of the final persistence or extinction outcome. We explain this by noticing that the first wave refers to the process of spatial invasion from the initially seeded patch, that can be quantified by the global invasion threshold~\cite{colizza_epidemic_2008}:
\begin{equation}
R_*=(R_0-1)\frac{\langle k^2\rangle -\langle k\rangle}{\langle k\rangle^2} \frac{2(R_0-1)^2}{\mu R_0^2}p \langle N\rangle\,,
\label{eq:R*}
\end{equation}
measuring the number of neighbouring patch that a seeded patch can infect. This is the analog of the basic reproductive number at the subpopulation level. Though this estimate was obtained for a SIR dynamics, we consider it here as we focus on the initial spatial invasion following the seeding, before the mechanism of immunity waning becomes important. The invasion condition $R_*>1$ indeed already proved to be a good approximation to define the upper bound of the low mobility regime in Eq.~(\ref{eq:gl_th}). Exploring different points in the intermediate mobility regime characterised by the same value of $p$ ($p=6\cdot 10^{-4}$ for panels $a$, $b$, $c$ and $h$), we find indeed the same initial invasion, regardless of the duration of the immunity period. 

Following the initial wave, dampened oscillations in the number of infected patches are observed at the spatial level, for large enough persistence probability. At the local level, oscillations in the incidence are caused  by the waning of immunity and have a frequency that decreases with $L$ and the patch population size, as predicted by the SIRS theory~\cite{pease_evolutionary_1987}. These local oscillations then interact at the spatial level because of the mobility coupling, translating in multiple waves in the spatial propagation. Those may be the result of traveling waves departing from well connected nodes 
~\cite{bjornstad_spatial_1999,grenfell_travelling_2001,marguta_impact_2015,grenfell_cities_1998}. Figure~\ref{fig:degree-block_analysis}$c$ indeed shows that if the pathogen is able to persist in the population, after the initial transient, the epidemic has more chances to persist in highly connected nodes. The more peripheral patches experience a sequence of sporadic outbreaks,  similarly to the low mobility case (panel $d$). Differently from before, here the local persistence is enhanced by the spatial coupling provided by the intermediate mobility. The continuous reseeding by infectious individuals traveling between subpopulations, in a situation of low level of synchronisation between local epidemics, fuel the circulation of the infection, hence allowing for a nonzero persistence even at high values of $L$, corresponding to the maximum discussed in Section~\ref{subsec:persistence}. 

This  {\it rescue effect} was previously introduced in the context of childhood diseases~\cite{bolker_impact_1996,bolker_space_1995,keeling_metapopulation_2000}. Here we find that it is driven by the increase in local persistence in highly connected nodes as illustrated in Figure~\ref{fig:persist_pop} and Figure~\ref{fig:degree-block_analysis}. In Figure~\ref{fig:persist_pop} the comparison of this mobility regime with the low mobility one shows that the rescue effect lowers the population threshold needed to have persistence $P_{gl} \geq 0.05$ and $P_{gl} \geq 0.95$. Moreover, comparing the intermediate and  low mobility regimes (blue curves in Figure~\ref{fig:degree-block_analysis}$a$ and $c$) for the case $L=150$, we see that while local persistence is not achieved when patches are fairly isolated, this is not the case for a large class of nodes as soon as mobility increases.

\subsubsection{High mobility coupling}
To characterize the unfolding of the epidemic in the high mobility regime we consider two points of panel $e$ of Figure~\ref{fig:dynamics} having the same traveling probability $p=10^{-2}$, and different lengths of the immunity period, namely $L=150$ and $L=250$, corresponding to $P_{gl}=1.0$ and $P_{gl}=0.60$, respectively. The epidemic dynamics in these cases show an oscillatory behaviour followed by an endemic equilibrium (Figure~\ref{fig:dynamics}$f$, $i$), similar to what was previously discussed in the intermediate mobility regime. Here frequencies of oscillations are however higher. This may be induced by an increase in synchronisation among patches, enabled by the stronger level of spatial coupling~\cite{hagenaars_spatial_2004}. Larger traveling fluxes lead indeed to almost all patches being infected during the first wave (e.g. $D/V\simeq 1$ for $p=10^{-2}$, panel $i$, compared to $D/V\simeq 0.6$ for $p=6\cdot 10^{-4}$, panel $h$, for the same value of $L$). Over time, these fluxes guarantee a large degree of seeding across patches along with increased mixing. For $L \lesssim 360$ and in a situation of global persistence, local persistence is found to be much higher across degree classes compared to intermediate mobility -- see e.g. curves in Figure~\ref{fig:degree-block_analysis} panel $e$ vs. panel $c$ for the same value of $L$ and consistently the values of $N_{min}$ in Figure~\ref{fig:persist_pop} corresponding to the $p=10^{-2}$ case. Spatial heterogeneities are therefore progressively reduced, with a larger number of patches behaving similarly and epidemic differences across patches almost disappearing. 

\begin{figure}[!ht]
\begin{center}
\includegraphics[width=0.65\linewidth]{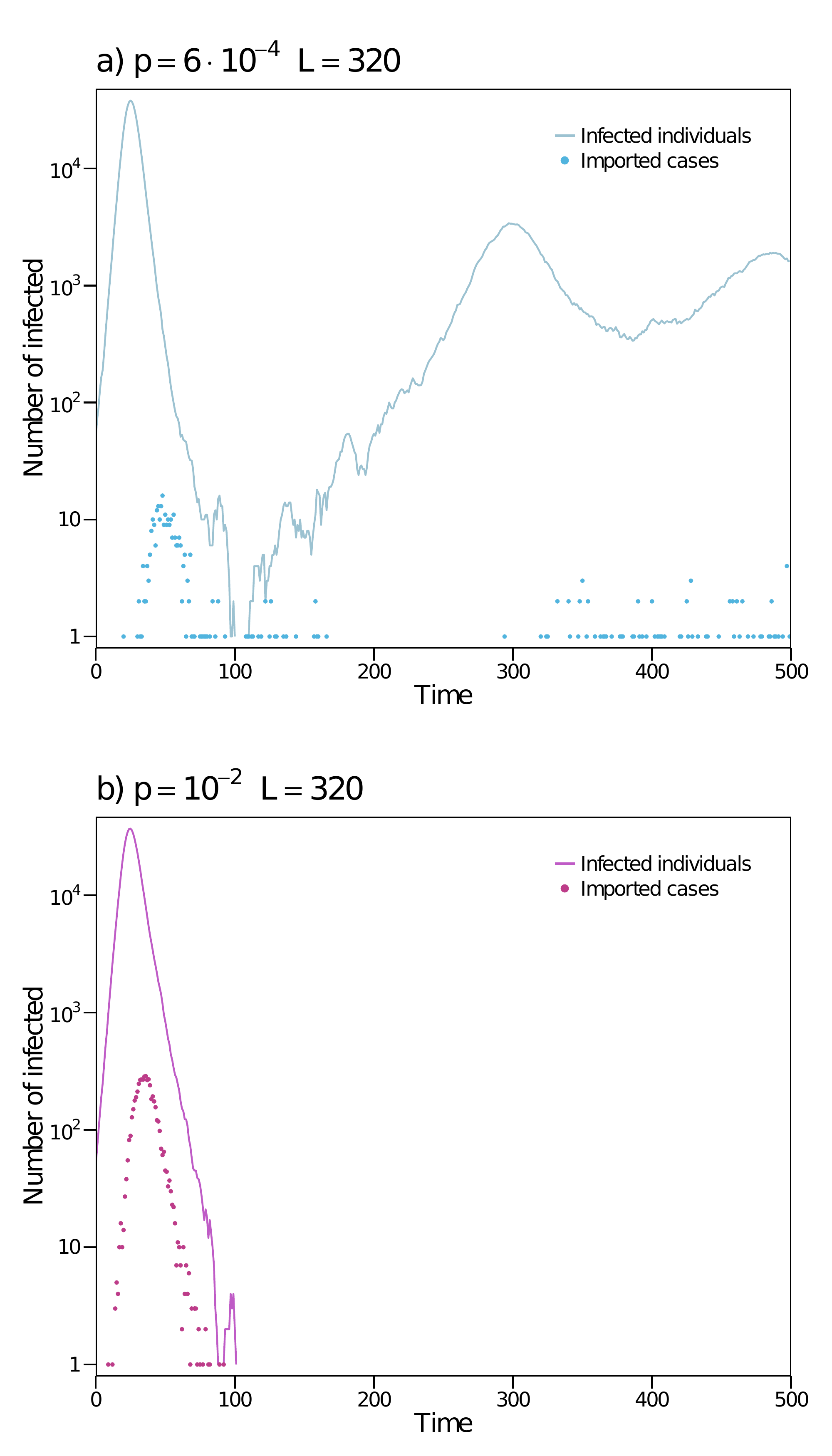}%
\caption{$(a)$ Temporal evolution of the fraction of infected individuals (curve) and imported cases (points) in the largest population of the metapopulation system ($k=91$, $N=2.5\cdot 10^5$) for $p=6\cdot 10^{-4}$ and duration of immunity $L=320$. $(b)$ Temporal evolution of the fraction of infected individuals (curve) and imported cases (points) in the same patch for $p=10^{-2}$ and duration of immunity $L=320$.}%
\label{fig:rescue_effect}%
\end{center}
\end{figure}

If persistence is enhanced locally for all patches, the resulting more synchronised system is however more prone to extinction of the epidemic for large enough immunity periods. This is the same type of phenomenon observed for measles in~\cite{bolker_impact_1996,bolker_space_1995,keeling_metapopulation_2000}, i.e. a decrease in spatial heterogeneity reduces the effectiveness of the rescue effect by enhancing the level of synchronization. To test this hypothesis, we compared the local and importation  dynamics occurring in the most connected patch in the intermediate and high mobility regimes during a single typical epidemic (Figure~\ref{fig:rescue_effect}). Immunity is maintained equal in both scenarios, for comparison. 
When $p=6\cdot 10^{-4}$ (intermediate mobility, panel $a$), importation of cases from neighbouring patches, desynchronised with respect to the local dynamics, allows the epidemic to survive the first wave and then quickly reach and endemic equilibrium, i.e. the rescue effect referred to before. If mobility is higher ($p=0.01$, panel $b$), importations are synchronised with the local dynamics, i.e. a peak of importations is occurring during the peak of the local epidemic wave, and both profiles fade out at the same time. In this way, there is no importation from outside the patch that could help sustain or relaunch the outbreak in the patch, and the epidemic goes extinct.

\section{Conclusion}

As it has been previously discussed in the literature, the epidemiological mechanism of waning of immunity alters the fate of an emerging infection and its chance to persist in the population. Here we have characterised in depth this effect once a spatially structured population with varying coupling is considered. By using a stochastic  individual-based metapopulation model that explicitly accounts for traveling fluxes among subpopulations, we have found that there are three distinct mobility regimes. Our results show that when mobility is low, the persistence conditions for the disease in the metapopulation system are the same as if the patches were isolated. When mobility is very high the epidemic behaves as if the whole metapopulation were well mixed. Eventually, at intermediate levels of mobility, the coupling between subpopulations creates rescue effects due to non-synchronised epidemics ongoing in the patches that maximise disease persistence. These effects are able to reseed the epidemics locally once it got extinct. 

Different persistence outcomes are associated to a variety of short-term dynamical behaviours. These are due to the interplay between local replenishment of susceptible individuals and the level of epidemic synchronisation across patches. For a very rapid waning of immunity, persistence occurs regardless of the level of mobility coupling, with local endemic equilibria allowing for the epidemic to propagate from one patch to another even at very low traveling probabilities. For longer durations of the immunity period, the probability of having local persistence is shaped by  the high heterogeneity in the population distribution (large patches may be above the critical size needed for local persistence, while small ones may not) and  the rescue effect. These two mechanisms are important at intermediate levels of traveling coupling and are responsible for the peak of persistence observed in this regime. 

Our findings are similar in many aspects to the ones obtained for the case of measles and, more in general, for a SIR model with vital dynamics and external introductions~\cite{bolker_impact_1996,lande_extinction_1998,keeling_metapopulation_2000,jesse_divide_2011,hagenaars_spatial_2004,grenfell_travelling_2001,marguta_impact_2015,grenfell_cities_1998,marguta_periodicity_2016,rohani_opposite_1999}. This exemplifies that different mechanisms of susceptible replenishment can lead to common dynamical features. Distinct and apparently contradicting behaviors have been shown in previous works~\cite{keeling_metapopulation_2000,rozhnova_impact_2014,jesse_divide_2011,hagenaars_spatial_2004}, highlighting the important role that mobility can have in triggering these mechanisms. Here, we have provided a detailed description of the epidemic dynamics for the case of a rapidly-spreading acute infection with waning of immunity, where the main statistical features of human space distribution and mobility are accounted for in a realistic way. We showed that the main result, i.e., that  persistence is maximised at intermediate mobility couplings, is valid for different levels of spatial fragmentation and also for a homogeneous distribution of population and connectivity.

In order to isolate the role of spatial structure and mobility we disregarded ingredients known to impact infectious disease dynamics in many real cases. In particular, we assumed no seasonality in transmission and no multi-strain interference. These two ingredients are known to alter the spread of diseases like influenza and respiratory syncytial virus, thus they are usually included in realistic data-driven modelling of these infections~\cite{zhang_co-circulation_2013,truscott_essential_2011,pitzer_environmental_2015}. Accounting for them here would largely increase the complexity of the dynamics, the number of parameters and the variety of possible outcomes in terms of persistence, transient dynamics and equilibrium state~\cite{poletto_host_2013,poletto_characterising_2015,roche_agent-based_2011,ballesteros_influenza_2009,rozhnova_impact_2014,buscarino_local_2014}, thus hindering a clear theoretical understanding. The chosen approximations may however limit the applicability of our work directly to real-case situations. For example, if we interpret the persistence diagram for the case of an emerging influenza strain~\cite{zhang_co-circulation_2013,boelle_transmission_2011}, we would obtain a 5\% probability for its persistence for $L \lesssim 3$ months in the low mobility regime, $L \lesssim 2$ years at the point of maximum of the intermediate mobility regime, $L \lesssim 1$ year in the high mobility regime. This contrasts with some estimates  for influenza up to $L\simeq 6$ years~\cite{truscott_essential_2011,bedford_integrating_2014,he_global_2015}). An open research question is therefore whether cross-immunity with previously circulating strains, seasonality and  environmental heterogeneity in transmission may interfere with space structure causing an increase in the threshold values of immunity duration. Starting from the complete characterisation of the epidemic dynamics proposed here, we plan to address these points in the future in order to better understand the epidemic dynamics resulting from a complex interplay of different factors.

\section*{Acknowledgments}
A.N.S.H. gratefully acknowledges L.H.G. Tizei for insightful discussions. S. M. acknowledges support from the Juan de la Cierva Program, Spain.
This work was partially supported by the EC-Health Contract PREDEMICS  (No. 278433) and the ANR Contract HARMSFLU (No. ANR-12-MONU-0018) to V.C.; the Government of Arag\'on, Spain through a grant to the group FENOL to Y.M.; the MINECO and FEDER funds (grant FIS2014-55867-P) to Y.M.; the European Commission FET-Proactive Project Multiplex (grant 317532) to Y.M.; the Ci\^encia sem Fronteiras programm - PDE/CNPq (No. 245936/2012-2) to A.N.S.H.

\bibliographystyle{naturemag}
\bibliography{persistence-SIRS_manuscript}
\end{document}